%% file: main_v3.tex
\documentclass[twocolumn,twocolappendix,times,tighten]{aastex631}

\usepackage[normalem]{ulem}  
\usepackage{bm}  
\usepackage{enumitem}  

\usepackage{stackengine}  

\usepackage{color}

\newcommand*{\mt}{\mathrm}

\newcommand{\figg}[1]{Fig.~\ref{fig:#1}}  

\newcommand*{\bp}{\beta_{\mathrm{pi1}}}

\newcommand*{\bpio}{\beta_{\mathrm{pi1}}}

\newcommand*{\kB}{k_{\mathrm{B}}}

\newcommand{\ms}{M_{s}}
\newcommand{\tiperp}{T_{\rm i\perp}}
\newcommand{\tipar}{T_{\rm i\parallel}}
\newcommand{\teperp}{T_{\rm e\perp}}
\newcommand{\tepar}{T_{\rm e\parallel}}

\newcommand*{\me}{m_{\mathrm{e}}}  
\newcommand*{\mi}{m_{\mathrm{i}}}

\newcommand*{\delgamio}{\theta_\mathrm{i1}}  
\newcommand*{\delgameo}{\theta_\mathrm{e1}}
\newcommand*{\Ti}{T_{\mathrm{i}}}
\newcommand*{\Teo}{T_{\mathrm{e1}}}  
\newcommand*{\Tio}{T_{\mathrm{i1}}}

\newcommand*{\Omci}{\Omega_{\mathrm{i}}}

\newcommand*{\rLio}{\rho_{\mathrm{i}}}

\newcommand*{\Ms}{M_{\mathrm{s}}}

\newcommand*{\Msim}{M_{\mathrm{0i}}}

\newcommand{\te}{T_{\rm e2}}
\newcommand{\tead}{T_{\rm e2,ad}}
\newcommand{\ti}{T_{\rm i2}}
\newcommand{\tiz}{T_{\rm i1}}
\newcommand{\tez}{T_{\rm e1}}
\newcommand{\tratio}{T_{\rm e1}/T_{\rm i1}}

\newcommand*{\htgequation}{\te/\tead-1\simeq  0.0016 M_s^{3.6}}

\received{\today}

\begin{document}

\title{Electron Heating in the Trans-Relativistic Perpendicular Shocks of Tilted Accretion Flows}

\author[0000-0002-1227-2754]{Lorenzo Sironi}
\affiliation{Department of Astronomy and Columbia Astrophysics Laboratory, Columbia University,
550 W 120th St.~MC~5246, New York, NY 10027, USA}
\affiliation{Center for Computational Astrophysics, Flatiron Institute, 162 5th Avenue, New York, NY, 10010, USA}
\email{lsironi@astro.columbia.edu}

\author[0000-0003-3483-4890]{Aaron Tran}
\affiliation{Department of Physics, University of Wisconsin--Madison, 1150 University Ave, Madison, WI 53706, USA}
\affiliation{Department of Astronomy and Columbia Astrophysics Laboratory, Columbia University,
550 W 120th St.~MC~5246, New York, NY 10027, USA}
\email{atran@physics.wisc.edu}

\begin{abstract}
    General relativistic magnetohydrodynamic (GRMHD) simulations of black hole tilted disks --- where the angular momentum of the accretion flow at large distances is misaligned with respect to the black hole spin --- commonly display standing shocks, within a few to tens of gravitational radii from the black hole. In GRMHD simulations of geometrically thick, optically thin accretion flows, applicable to low-luminosity sources like Sgr A* and M87*, the shocks have trans-relativistic speed, moderate plasma beta (the ratio of ion thermal pressure to magnetic pressure is $\bp\sim 1-8$), and low sonic Mach number (the ratio of shock speed to sound speed is $M_s\sim 1-5$). We study such shocks with two-dimensional particle-in-cell simulations and we quantify the efficiency and mechanisms of electron heating, for the special case of pre-shock magnetic fields perpendicular to the shock direction of propagation. We find that the post-shock electron temperature $T_{\rm e2}$ exceeds the adiabatic expectation $T_{\rm e2, ad}$ by an amount $\htgequation$, nearly independent of the plasma beta and of the pre-shock electron-to-ion temperature ratio $\tez/\tiz$, which we vary from $0.1$ to unity. We investigate the heating physics for $M_s\sim 5-6$ and find that electron super-adiabatic heating is governed by magnetic pumping at $\tez/\tiz=1$, whereas heating by $B-$parallel electric fields (i.e., parallel to the local magnetic field) dominates at $\tez/\tiz=0.1$. Our results provide physically-motivated subgrid prescriptions for electron heating at the collisionless shocks seen in GRMHD simulations of black hole accretion flows.
\end{abstract}
\keywords{
Galaxy accretion disks (562), Stellar accretion disks (1579), Shocks (2086), Plasma astrophysics (1261)
}

\section{Introduction}

Electrons emit the light we see from accreting black holes, including the famed Event
Horizon Telescope (EHT) images of M87* and Sagittarius A* (Sgr A*)
\citep{eht2019-m87-i,eht2022-sgra-i}. Yet, the electron temperature in such systems, and hence the source of their luminosity, is uncertain. In low-luminosity sources like Sgr A* and M87*, the density in the hot, geometrically-thick accretion flow is so low that the plasma is nearly collisionless. Therefore, wave-particle interactions regulate the energy exchange between protons and electrons. In recent years, analytical models and plasma simulations have been used to study the efficiency of electron heating, in case energy dissipation is governed by magnetic reconnection \citep{rowan_17,rowan_19} or plasma turbulence \citep[e.g.,][]{howes_10,zhdankin_19,zhdankin_21,kawazura_19,kawazura_20,comisso_22,squire_23,arza_19,arza_23}. Physically-motivated inputs for the electron heating rate can then be incorporated into general relativistic magnetohydrodynamic (GRMHD) simulations and used to produce synthetic images and spectra to compare with  observations.

In recent years, GRMHD simulations of ``tilted'' disks --- where the angular momentum of the accretion flow at large distances is misaligned with respect to the black hole spin --- have shown that shocks form within a few to tens of gravitational radii from the black hole \citep{fragile2001,fragile2007,fragile_08,mckinney_13,zhuravlev_14,morales_14, dexter_11,dexter_13,white_19,white2020-tilt,white2022,bollimpalli_22,bollimpalli_23, tsokaros_22,musoke_23,liska_23,chatterjee_23,ressler_23,kaaz_23}, in agreement with earlier analytical arguments \citep[e.g.,][]{ogilvie_99,ogilvie_latter_13,fairbairn_21}.
Tilted disks are of general interest  because (1) the accretion disk around Sgr A* could
be tilted within EHT constraints \citep{eht2022-sgra-v},
and (2) dynamics within tilted disks may help explain the time-varying emission
from Sgr A* or the mysterious quasi-periodic oscillations (QPOs) of galactic X-ray binaries (XRBs).
In weakly collisional tilted disks (as well as in aligned disks, see \citealt{conroy_23}), shocks then offer a novel channel for energy dissipation and electron heating --- in addition to reconnection and turbulence.
It is therefore timely to assess if, and how much, proton energy can be
transferred to electrons at collisionless shocks, for the conditions expected in tilted accretion flows.

In this paper, we use two-dimensional (2D) particle-in-cell (PIC) simulations to quantify the efficiency and mechanisms of electron heating, for the special case of pre-shock magnetic fields perpendicular to the shock direction of propagation. We are primarily motivated by the shock conditions extracted by \citet{generozov2014} from the GRMHD simulation by \citet{fragile2007} of a radiatively inefficient, geometrically thick accretion flow.
These shocks have trans-relativistic speed (the shock-frame upstream Lorentz factor is $\sim 1.2-1.8$), moderate ion beta $\beta_{\rm pi1}\sim 1-8$ (the ratio of ion pressure to magnetic pressure), and low sonic Mach number $M_s\sim 1-5$ (the ratio of shock speed to sound speed). Both the shock velocity and the Mach number increase for larger tilt angles (compare Figs.~6 and 7 in \citealt{generozov2014}).
While extensive literature exists on electron heating in non-relativistic shocks \citep[e.g.,][]{raymond_23}, the plasma conditions most relevant for collisionless shocks in tilted accretion disks are still unexplored. The regime of low sonic Mach number and moderate-to-high plasma beta is similar to the case of merger shocks in galaxy clusters studied by \citet{guo_17,guo_18}, yet the flow velocity in black hole disks is much faster than in the intracluster medium, and the study of such shocks deserves a separate investigation.

This paper is organized as follows.
We describe the setup of our PIC simulations in Section \ref{sec:methods}, and present the general structure of the shocks in Section \ref{sec:results}. In Section \ref{sec:heating}, we discuss the physics of electron heating, and show that the post-shock electron temperature $T_{\rm e2}$ exceeds the adiabatic expectation $T_{\rm e2, ad}$ by approximately $\htgequation$, nearly independent of the plasma beta and of the pre-shock ion-to-electron temperature ratio $\tez/\tiz$, which we vary from 0.1 to unity. As we discuss in Section \ref{sec:discuss}, this fitting formula can be used to incorporate the electron shock-heating physics into GRMHD simulations of tilted accretion disks.

\section{Simulation Setup} \label{sec:methods}

We simulate 2D ion-electron shocks using the relativistic particle-in-cell
(PIC) code TRISTAN-MP \citep{buneman1993, spitkovsky2005}.
Our shocks are formed by reflecting a left-ward traveling flow off a stationary
wall at $x=0$; the shock travels from left to right along $+\hat{x}$.
The simulation (lab) frame is the downstream rest frame.
Plasma is injected from the right-side $x$ boundary, which continuously recedes from the wall
to remain ahead of the shock at all times.
The $y$ boundary is periodic.

Subscript $0$ refers to upstream quantities measured in the simulation frame.
Subscript $1$ refers to upstream quantities measured in the upstream rest frame.
Subscript $2$ refers to downstream quantities measured in the downstream rest frame (which coincides with the simulation frame).
An exception is made for the 3-velocities $v_1, v_2$ and the 4-velocities
$u_1,u_2$ (where $u_1=v_1/\sqrt{1-(v_1/c)^2}$, and similarly for $u_2$), which are measured in the shock frame.

The upstream flow is a drifting ion-electron plasma with 3-velocity $v_0$
(Lorentz factor $\gamma_0=1/\sqrt{1-(v_0/c)^2}$), single-species density $n_0$, and magnetic field
$B_0$ in the simulation frame.
The upstream magnetic field has an angle $\theta_{Bn0}=90^\circ$ with respect to the $\hat{x}$ direction of the shock normal, and it lies along the $y$ direction (in \citealt{guo_17}, we demonstrated that this in-plane geometry is most suitable for studying electron heating in low Mach number shocks, as compared to the alternative case of out-of-plane fields oriented along $z$).
Ions are singly-charged and the plasma is charge neutral. We employ the realistic mass ratio $\mi/\me=1836$.
The rest-frame single-species upstream density is $n_1 = n_0/\gamma_0$.
Both ions and electrons are Maxwell-J\"{u}ttner distributed with initial
temperatures $\Tio$ and $\Teo$ respectively.
The dimensionless temperature is $\theta_\mt{s1} = \kB T_\mt{s1}/(m_\mt{s1} c^2)$,
where subscript $s \in \{\mt{i},\mt{e}\}$ indicates particle species.

The relative balance of rest-mass, thermal, magnetic, and kinetic energies in the upstream plasma is
fully specified by dimensionless ratios.
The ion dimensionless temperature $\theta_{\rm i1}$ specifies the relative balance of thermal and rest-mass energy. Motivated by GRMHD simulations, we fix $\theta_{\rm i1}=0.01$.  The upstream ion plasma beta $\beta_\mt{pi1} = 8\pi P_{\rm i1}/B_1^2$ is the ratio between the ion thermal pressure and the magnetic pressure, which we vary in the range $1\leq \beta_\mt{pi1}\leq 8$. The ratio between kinetic and thermal energies is set by the sonic Mach number $\Ms = v_1/c_{s1}$,
where the upstream sound speed
${c_{s1}} = \sqrt{(\Gamma_{\rm i} \delgamio + \Gamma_{\rm e} \delgameo \me/\mi) c^2 / h}$
with $\Gamma_{\rm i} = 5/3$, $\Gamma_{\rm e} = 4/3$, and specific enthalpy
$h \approx 1 + 5\delgamio/2 + 4\delgameo \me/\mi$ for non-relativistic ions
and relativistic electrons
(equivalently, one could use the Alfv\'enic Mach number
or the magnetosonic Mach number).
Since we set up our simulation in the downstream rest-frame, we
cannot choose $M_s$ directly; instead, we control the simulation-frame,
ion-sound Mach number $\Msim = v_0 / c_{\rm si1}$ with
$c_{\rm si1} = \sqrt{\Gamma_{\rm i} \delgamio c^2}$, and we measure $M_s$ after the
simulation ends.
We explore the dependence of electron heating on
$\Msim$, which we vary from 2 to 5. Finally,
in the absence of efficient collisional coupling, ions and
electrons might have different temperatures ahead of the
shock, so we vary $\tez/\tiz$ from 0.1 to 1. It follows that the
dimensionless electron temperature in the upstream varies
in the range $\delgameo = 1.84$--$18.4$.
The resulting sonic Mach number $M_s$ varies from 2.6 to 6.1.
For $\tez/\tiz=1$, the ratio $M_s/\Msim \sim 0.9$ to $1.3$;
for $\tez/\tiz=0.1$, the ratio $M_s/\Msim \sim 1.2$ to $1.6$.

We define reference plasma scales and parameters based on the upstream flow
properties. We initialize the pre-shock medium with 16 particles per cell per species in the simulation frame.
The plasma frequency is $\omega_\mt{ps} = \sqrt{4\pi n_1 e^2 / m_\mt{s}}$
and the plasma skin depth is $d_\mt{s} = c/\omega_\mt{ps}$. The transverse width of the domain in the $y$ direction is $22.4\,d_\mt{i}$.
We resolve the electron skin depth $d_\mt{e}$ with 3 cells. For our choice of $\theta_{\rm i1}=0.01$ and $\tez/\tiz\geq 0.1$, the electron dimensionless temperature is $\theta_{\rm e1}\geq 1.8$, so the electron Debye length $\lambda_\mt{De} = \sqrt{\kB \Teo/(4\pi n_1 e^2)}$ is always well resolved.
We measure time in units of the inverse ion cyclotron frequency defined with lab-frame quantities, $\Omega_\mt{i} = e B_0/(m_\mt{i} c)$, and length in units of the ion Larmor radius $\rLio = {\gamma_0 v_0 \mi c}/(e B_0)$ (still defined with lab-frame quantities). It is not obvious whether lab-frame quantities are the most appropriate to use in our definitions of time and length units. Nevertheless, our definitions suffice up to
order-unity corrections.

We compute the Mach number $M_s$ as follows. At the end of the simulations ($\Omci t \sim 25$ for all cases apart from $\bpio=1$, where we evolve until $\Omci t \sim 40$), we identify the shock position $x_\mt{shock}$ as the right-most ion
density peak.
We then estimate the shock-frame flow velocities as
$v_2 = x_\mt{shock}/t$ and $v_1 = (v_0 + v_2) / (1 - v_0v_2/c^2)$, which yield
a measurement of $M_s = v_1/c_{\rm s1}$.

A complete list of the input parameters of our simulations is in  Table~\ref{tab:param} of the Appendix.

\begin{figure}[!h]
    \centering
    \includegraphics[width=3.375in]{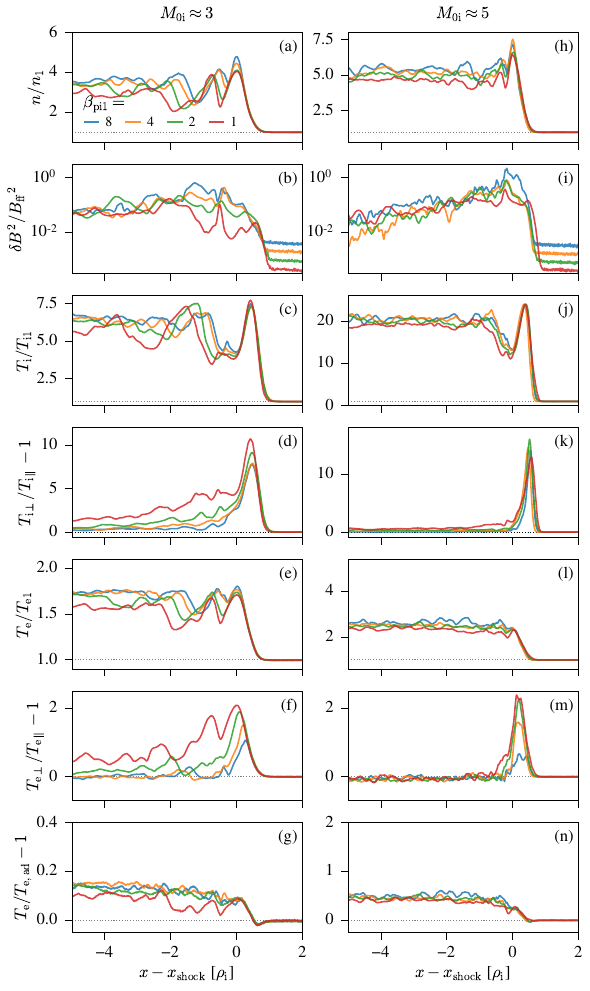}
    \caption{Dependence on $\Msim$ and $\bp$ of various $y$-averaged quantities measured at $\Omega_{\rm i}t\sim 25$ (with the exception of $\bpio=1$, which is measured at $\Omega_{\rm i}t\sim 40$), for a pre-shock temperature ratio $\tez/\tiz=1$. The $x$ coordinate is measured relative to the shock location, in units of the proton Larmor radius $\rLio$. From top to bottom, we plot: (a) rest-frame number density; (b) energy in magnetic fluctuations, normalized to the energy of the frozen-in  field (see text); (c) mean proton temperature (see text); (d) proton temperature anisotropy; (e) mean electron temperature; (f) electron temperature anisotropy; (g) excess of electron temperature beyond the adiabatic prediction for an isotropic 3D ultra-relativistic gas. Note that the vertical axis range is different between left and right columns.}
    \label{fig:avg1d-vary-bpi-teti1e0}
\end{figure}

\begin{figure}[!h]
    \centering
    \includegraphics[width=3.375in]{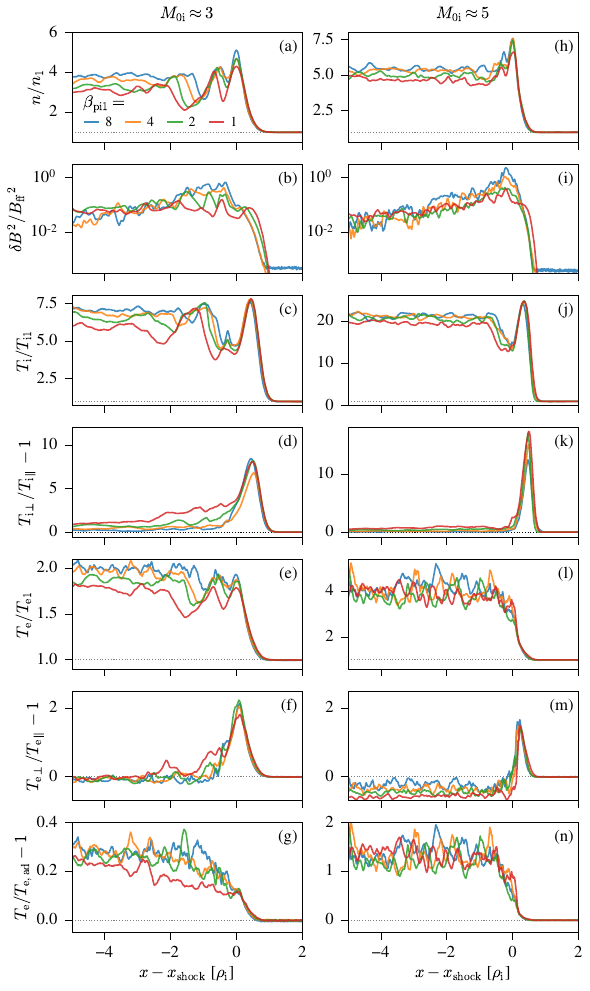}
    \caption{
        Like Figure~\ref{fig:avg1d-vary-bpi-teti1e0}, but for $T_\mt{e1}/T_\mt{i1} = 0.1$.
    }
    \label{fig:avg1d-vary-bpi-teti1e-1}
\end{figure}

\section{Shock Structure}
\label{sec:results}
\begin{figure*}
    \centering
    \includegraphics[width=\textwidth]{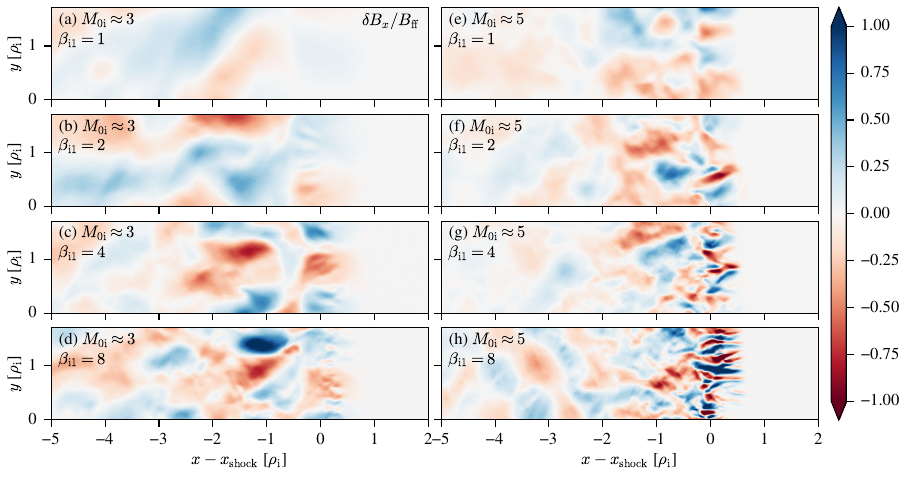}
    \caption{
        Dependence on $\Msim$ and $\bp$ of the fluctuating magnetic field component $\delta B_x/B_{\rm ff}$ measured at $\Omega_{\rm i}t\sim 25$ (for $\bpio=1$, $\Omega_{\rm i}t\sim 40$), assuming $\tez/\tiz=1$. The field is measured in the simulation frame, and the $x$ coordinate is measured relative to the shock location.
    }
    \label{fig:bx-vary-bpi-teti1e0}
\end{figure*}

\begin{figure*}
    \centering
    \includegraphics[width=\textwidth]{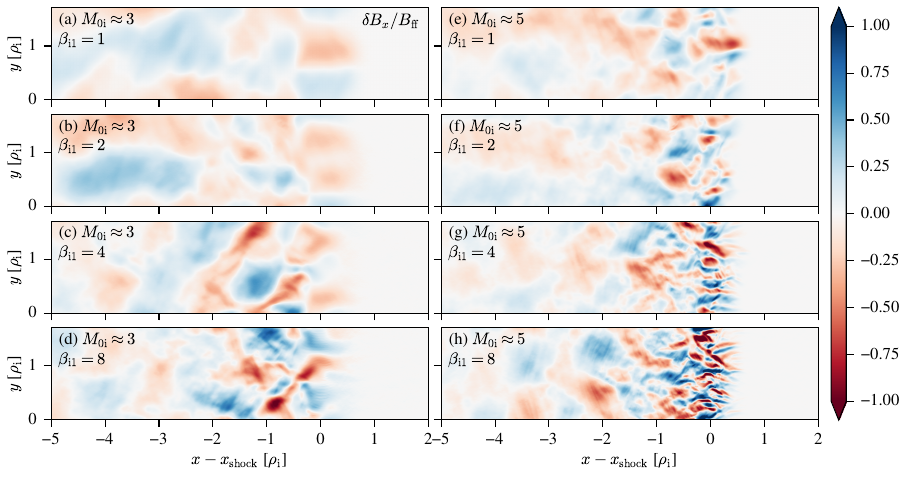}
    \caption{
    Like Figure~\ref{fig:bx-vary-bpi-teti1e0}, but for $T_\mt{e1}/T_\mt{i1} = 0.1$.
    }
    \label{fig:bx-vary-bpi-teti1e-1}
\end{figure*}

The dependence of the shock structure on the upstream conditions is illustrated in Figs.~\ref{fig:avg1d-vary-bpi-teti1e0}-\ref{fig:bx-vary-bpi-teti1e-1}. Figs.~\ref{fig:avg1d-vary-bpi-teti1e0} and \ref{fig:avg1d-vary-bpi-teti1e-1} show $y$-averaged quantities, as a function of the Mach number (in each figure, $\Msim\approx3$ in the left column and $\Msim\approx5$ in the right column), the ion plasma beta (different colors in each plot, see legend in panel (a)), and the electron-to-ion temperature ratio ($\tez/\tiz=1$ in \figg{avg1d-vary-bpi-teti1e0} and $\tez/\tiz=0.1$ in \figg{avg1d-vary-bpi-teti1e-1}). We first discuss the dependence on the Mach number and the ion plasma beta, and then on the electron-to-proton temperature ratio.

In agreement with the Rankine-Hugoniot relations, the ion density jump is larger for higher $\Msim$ (compare panels (a) and (h) in \figg{avg1d-vary-bpi-teti1e0} and \figg{avg1d-vary-bpi-teti1e-1}). As regard to the dependence on $\bp$, it is rather modest, with only marginal evidence for weaker compressions in the most magnetized case of $\bp=1$.
As a result of flux freezing alone, one would expect the lab-frame magnetic field to be $B_{\rm ff}=(\langle n\rangle_y/n_0) B_0$, where
$\langle \cdot \rangle_y$ denotes averaging along the $y$ direction. In reality, the magnetic field energy at the shock and in the downstream region exceeds the expectation from flux freezing, due to self-generated magnetic fluctuations. Their strength is quantified by $\delta B^2=B_x^2+(B_y-B_{\rm ff})^2+B_z^2$ in panels (b) and (i) of \figg{avg1d-vary-bpi-teti1e0} and \figg{avg1d-vary-bpi-teti1e-1}. As we further discuss below, the relaxation of ion velocity-space anisotropies can result in proton cyclotron modes and mirror modes (for a review of anisotropy instabilities in relativistic plasmas, see \citealt{alisa_23}). For the magnetic geometry employed in this paper, proton cyclotron waves would appear in  $B_x$ and $B_z$, and their wavevector is aligned with the mean field; in contrast, mirror modes appear in $B_x$ and $B_y$, and their wavevector is oblique with respect to the mean field. In \figg{bx-vary-bpi-teti1e0} and \figg{bx-vary-bpi-teti1e-1}, we show the $x$ component $\delta B_x/B_{\rm ff}=B_x/B_{\rm ff}$, which includes both proton cyclotron and mirror modes.
We find that proton cyclotron modes dominate near the shock.

We define the isotropic-equivalent proton temperature $T_{\rm i}=(2\tiperp+\tipar)/3$,
which we present in panels (c) and (j) of Figs.~\ref{fig:avg1d-vary-bpi-teti1e0} and \ref{fig:avg1d-vary-bpi-teti1e-1}. We define $\tiperp$ as the proton temperature perpendicular to the mean field, and $\tipar$ as the proton temperature along the mean field. It is apparent that $T_{\rm i}/\tiz$ increases with $\Msim$, which comes from the fact that the temperature jump predicted by the Rankine-Hugoniot relations for the overall fluid is a monotonic function of $M_s \sim \Msim$, and that most of the post-shock fluid energy resides in protons (rather than electrons or proton-driven waves).

At the shock, magnetic fluctuations are sourced by the relaxation of the proton temperature anisotropy $\tiperp/\tipar$ (panels (d) and (k) in \figg{avg1d-vary-bpi-teti1e0} and \figg{avg1d-vary-bpi-teti1e-1}), which is larger for higher $\Msim$. This has two consequences: (\textit{i}) the
 greater amount of free energy stored in proton temperature anisotropy for higher $\Msim$ generates stronger waves (compare  panels (b) and (i) in Figs.~\ref{fig:avg1d-vary-bpi-teti1e0}-\ref{fig:avg1d-vary-bpi-teti1e-1}); (\textit{ii}) linear theory prescribes that the waves grow faster for higher levels of anisotropy (so, higher $\Msim$). In fact,  panels (b) and (i) in Figs.~\ref{fig:avg1d-vary-bpi-teti1e0} and \ref{fig:avg1d-vary-bpi-teti1e-1} show that the peak of wave activity is located right at the shock for $\Msim\approx5$, but shifts farther downstream for lower $\Msim$, due to the slower wave growth. As regard to the dependence on $\bp$, we find that the proton anisotropy at the shock is nearly insensitive to $\bp$. However, proton-generated waves are  stronger for higher $\bp$, when normalized to the flux-frozen field (see panels (b) and (i) in Figs.~\ref{fig:avg1d-vary-bpi-teti1e0}-\ref{fig:avg1d-vary-bpi-teti1e-1}, as well as Figs.~\ref{fig:bx-vary-bpi-teti1e0} and \ref{fig:bx-vary-bpi-teti1e-1}). This is because the free energy in proton anisotropy available to source the waves is larger for higher $\bp$, when compared to the magnetic energy of the background field.

 Due to pitch angle scattering by the proton modes, the proton anisotropy drops behind the shock at a faster rate for higher $\Msim$ and higher $\bp$, since the waves grow faster and are stronger. Far downstream, the proton anisotropy is expected to be reduced below a marginal stability threshold, which is lower at higher plasma beta for both mirror and proton cyclotron modes. A decrease in anisotropy with increasing $\bp$ is apparent in panels (d) and (k) of Figs.~\ref{fig:avg1d-vary-bpi-teti1e0} and \ref{fig:avg1d-vary-bpi-teti1e-1}, especially at low $\Msim$. It is worth noting that low-$\bp$ low-$\Msim$ shocks maintain an appreciable degree of proton anisotropy in the far downstream, so the resulting adiabatic index will be larger than for a 3D isotropic gas. Then the plasma will be less compressible, which explains why the red curve in the density profile of panels (a) and (h) lies below the other lines.

So far, we have focused on the proton physics. As regard to electrons, we find that the isotropic-equivalent post-shock electron temperature $T_{\rm e}=(2\teperp+\tepar)/3$ increases for greater $\Msim$ (compare panels (e) and (l) in Figs.~\ref{fig:avg1d-vary-bpi-teti1e0} and \ref{fig:avg1d-vary-bpi-teti1e-1}). This might just follow from the dependence on $\Msim$ of the adiabatic heating efficiency, since the density compression increases with $\Msim$. However, the efficiency of irreversible electron heating is also higher at larger $\Msim$. In panels (g) and (n), we present the excess of electron temperature beyond the adiabatic expectation $T_{{\rm e, ad}}=(n/n_1)^{1/3}\tez$ appropriate for a 3D isotropic ultra-relativistic gas. The assumption of isotropic electrons is well justified in the downstream region, where $\teperp\simeq \tepar$ (panels (f) and (m) in Figs.~\ref{fig:avg1d-vary-bpi-teti1e0} and \ref{fig:avg1d-vary-bpi-teti1e-1}).

A large fraction of the electron irreversible heating comes from magnetic pumping \citep{hollweg_85,Berger1958,Borovsky1986,guo_17,ley_23}. In this mechanism, two  ingredients are needed: (\textit{i}) the presence of an electron temperature anisotropy, which in our case is induced by field amplification coupled to adiabatic invariance; and (\textit{ii}) a mechanism to break the electron adiabatic invariance. Field amplification in our shocks has two potential drivers: at the shock ramp, density compression coupled to flux freezing leads to field amplification; in addition, at the shock front and further downstream, proton waves accompanying the relaxation of the proton temperature anisotropy contribute to further field growth. As regard to the mechanism for breaking the electron adiabatic invariance, in non-relativistic low-$\ms$ and high-$\bp$ shocks it was attributed to pitch angle scattering by whistler waves sourced by the electron anisotropy itself \citep{guo_17,guo_18,ha_21,ha_21b,ley_24}. For the trans-relativistic conditions of this work ($\theta_{\rm i}\lesssim 1$ and $\theta_{\rm e}\gg1$), the ratio between proton and electron Larmor radii (which roughly corresponds to the ratio of proton cyclotron wavelength to whistler wavelength) is $\sim (T_{\rm i}/T_{\rm e})/\sqrt{\theta_{\rm i}}$. At the shock $\theta_{\rm i}\sim 0.1$ and $T_{\rm i}/T_{\rm e}$ is a few times larger than $T_{\rm i1}/T_{\rm e1}$ (see Figs.~\ref{fig:avg1d-vary-bpi-teti1e0} and \ref{fig:avg1d-vary-bpi-teti1e-1}). This implies that the proton cyclotron wavelength is larger than the whistler wavelength, but their ratio is smaller than for non-relativistic temperatures, where it is $\sim \sqrt{m_{\rm i}T_{\rm i}/(m_{\rm e}T_{\rm e})}\gg1$.
The presence of short-wavelength electron whistler waves is mostly supported by the $\Msim\approx 3$ cases (\figg{bx-vary-bpi-teti1e0}(b)-(d) and \figg{bx-vary-bpi-teti1e-1}(d)). For $\Msim\approx5$, proton-driven modes grow quickly and reach strong amplitudes. They dominate the wave energy at the shock, hiding the potential presence of whistler waves.

The amount of super-adiabatic electron heating is nearly independent of $\bp$, with the exception of $\bp=1$ in the $\Msim\approx3$ shock (red line in panel (g) of Figs.~\ref{fig:avg1d-vary-bpi-teti1e0} and \ref{fig:avg1d-vary-bpi-teti1e-1}). This case displays the lowest density compression and the weakest level of proton-driven waves (see panel (b) in the same figures), so it lacks a sufficient degree of field amplification to drive efficient super-adiabatic electron heating via the pumping mechanism. In contrast, electron heating beyond the adiabatic expectation is a strong function of $\Msim$. First of all, the electron fluid suffers a stronger compression while passing through the ramp of a higher-$\Msim$ shock \citep{guo_17,guo_18}. In addition, the highly anisotropic protons present in higher-$\Msim$ shocks generate stronger proton modes. In both cases, stronger field amplification at higher-$\Msim$ shocks performs more work on the electrons and ultimately leads to greater electron heating.

By comparing Figs.~\ref{fig:avg1d-vary-bpi-teti1e0} and \ref{fig:avg1d-vary-bpi-teti1e-1} (panels (g) and (n)), we infer that the amount of super-adiabatic heating is larger for $\tratio=0.1$ than for $\tratio=1$, by roughly a factor of two. As we further discuss in Section \ref{sec:heating}, this trend cannot be explained by the magnetic pumping framework discussed so far. In fact, both the amount of field amplification (panels (b) and (i)) as well as the degree of electron anisotropy (panels (f) and (m)) are nearly insensitive to  $\tratio$, at fixed $\bp$ and $\Msim$. Thus, we would expect comparable amounts of pumping-driven heating for $\tratio=0.1$ and $\tratio=1$ (we will confirm in Section \ref{sec:heating} that this is indeed the case). Below, we demonstrate that the greater heating efficiency of $\tratio=0.1$ shocks is due to the dominant contribution of $B-$parallel electric fields (i.e., $E_\parallel= \boldsymbol{E}\cdot \boldsymbol{B}/B$). Heating by $E_\parallel$ tends to increase $\tepar$, which explains why $\teperp<\tepar$ in panel (m) of \figg{avg1d-vary-bpi-teti1e-1}.

\begin{figure}[!t]
    \centering
    \includegraphics[width=\columnwidth]{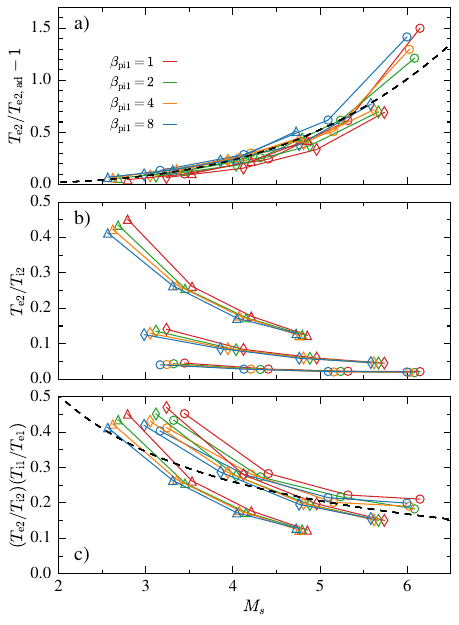}
    \caption{
Amount of super-adiabatic electron heating (panel (a)) and post-shock electron-to-ion temperature ratio (panels (b) and (c)), as a function of $\ms$ (horizontal axis), $\bp$ (colors, see legend) and $\tratio$ (triangles for $\tratio=1$, diamonds for $\tratio=0.3$, circles for $\tratio=0.1$). Panels (a) and (b) present our raw data, while panel (c) condenses the dependence on pre-shock parameters in a simpler form.
    }
    \label{fig:eff}
\end{figure}

\section{Electron Heating Efficiency and Mechanism} \label{sec:heating}
We now characterize the efficiency of electron heating in our shocks as a function of the proper Mach number $M_s$. We measure the particle density $n_2$ and the isotropic-equivalent temperatures $\te$ and $\ti$ in a region that is sufficiently far downstream that the temperatures have reached a quasi-steady value (see Table~\ref{tab:param}). The post-shock electron temperature exceeds the adiabatic expectation $\tead=(n_2/n_1)^{1/3}\tez$ by the amount indicated in \figg{eff}(a). There, different colors indicate different $\bp$ (see the legend), while different symbols specify the value of $\tratio$: triangles for $\tratio=1$, diamonds for $\tratio=0.3$, circles for $\tratio=0.1$. The amount of super-adiabatic heating is nearly independent from $\bp$ and $T_{\rm e1}/T_{\rm i1}$, and it is an increasing function of $M_s$. Its dependence on $M_s$ can be parameterized as:
\begin{equation}
\htgequation
\end{equation}
as indicated by the dashed line.

We also present the dependence on $\bp$, $\ms$ and $\tratio$ of the post-shock electron-to-ion temperature ratio $\te/\ti$ in \figg{eff}(b). The dependence on $\bp$ is  weak, while the dependence on $\ms$ and $\tratio$  can be approximately cast as
$(\te/\ti)(T_{\rm i1}/T_{\rm e1})\simeq \ms^{-0.8}-0.07$ (dashed line in \figg{eff}(c)).
In all cases $\te/\ti< \tratio$, i.e., shocks systematically lead to temperature disequilibration.

\begin{figure}[!t]
    \centering
    \includegraphics[width=0.5\textwidth]{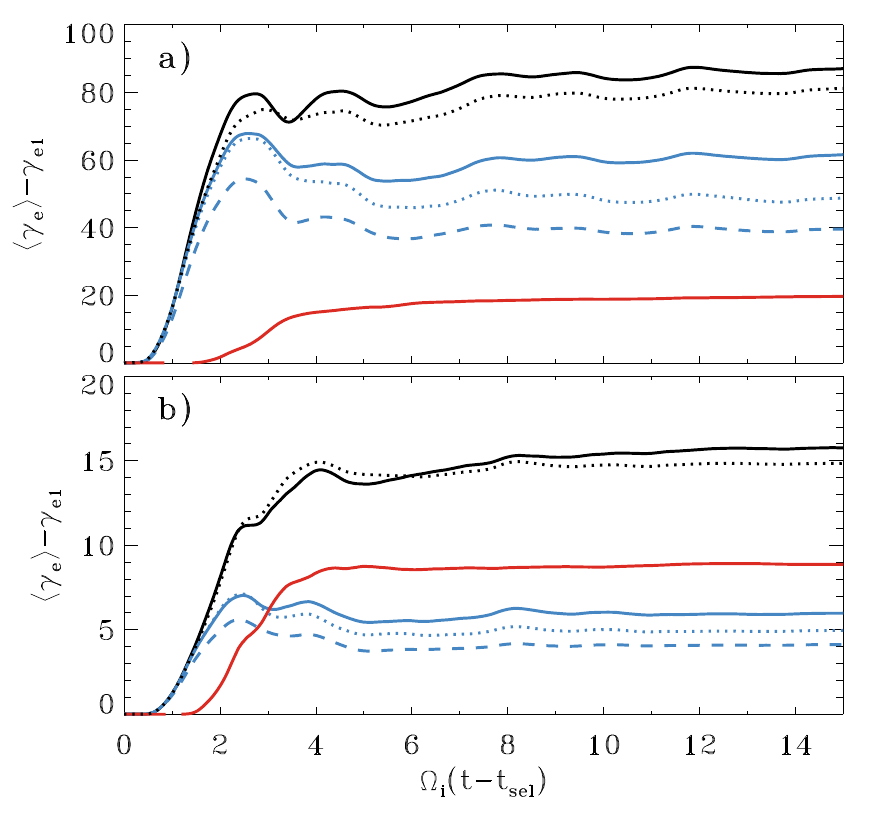}
    \caption{
Time evolution of the mean energy of a population of electrons tracked during their passage through the shock, as measured in the local $E\times B$ frame. We fix $\Msim=5$ and $\bp=2$ and explore two cases: $\tratio=1$ (top) and $\tratio=0.1$ (bottom). The black solid line indicates $\langle \gamma_{\rm e} \rangle-\gamma_{\rm e1}$; the red line illustrates the work done by $B-$parallel electric fields; the blue solid line indicates heating by magnetic compression, while the dashed and dotted blue lines correspond to the adiabatic expectations for a 3D and 2D gas, respectively (see text for details); super-adiabatic heating via magnetic pumping is the difference between solid and dashed blue lines; the dotted black line is the sum of the red and blue solid lines.
    }
    \label{fig:line}
\end{figure}

\begin{figure}
    \centering
    \includegraphics[width=0.5\textwidth]{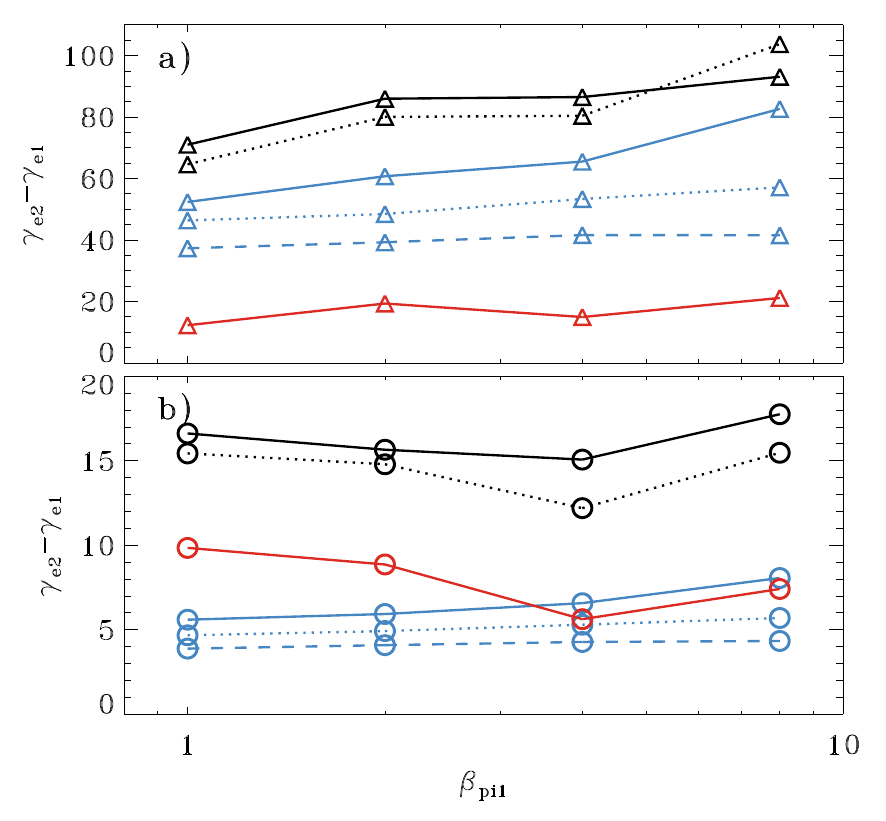}
    \caption{Contributions of various heating mechanisms to the mean energy change of the tracked electrons, as measured in the local $E\times B$ frame.  We fix $\Msim=5$ and explore the dependence on $\bp$ (horizontal axis) and $\tratio$ ($\tratio=1$ at the top and $\tratio=0.1$  at the bottom).
    The data points are obtained by time-averaging the heating curves (e.g., the ones in \figg{line}) at $\Omega_{\rm i}(t-t_{\rm sel})\gtrsim 8$. The color coding and the line style correspond to \figg{line}: the black solid points indicate $\gamma_{\rm e2}-\gamma_{\rm e1}$; the red points illustrate the work done by $B-$parallel electric fields; the blue points connected by solid lines indicate heating by magnetic compression, while the dashed and dotted blue lines correspond to the adiabatic expectations for a 3D and 2D gas, respectively; super-adiabatic heating via magnetic pumping is the difference between solid and dashed blue lines; the black points connected by dotted black lines are the sum of magnetic compression and $B-$parallel heating.}
    \label{fig:all}
\end{figure}

In Figs.~\ref{fig:line} and \ref{fig:all}, we consider shocks with $\Msim=5$ and investigate the dominant mechanisms of electron heating. In \figg{line}, we also fix $\bp=2$ and compare two cases: $\tratio=1$ (top) and $\tratio=0.1$ (bottom). At time $t_{\rm sel}$ (where $\Omega_{\rm i}t_{\rm sel}\sim 40$ for $\bp=1$ and $\Omega_{\rm i}t_{\rm sel}\sim 25$ for all other cases), we select a slab of electrons just upstream of the shock foot, having roughly the same initial $x$ location (within 10\%). Each simulation selects approximately 0.5 million electrons.
We  follow them in time until they propagate  far enough behind the shock that their mean energy approaches roughly a constant value. Their properties are recorded with an output cadence of 50 timesteps $=7.5/\omega_{\rm pe}$.
In \figg{line}, we define $\gamma_{\rm e1}\simeq 3\,\theta_{\rm e1}$ as their initial mean Lorentz factor, while $\langle\gamma_{\rm e}\rangle$ (black solid lines in \figg{line}) is, at any subsequent time, the mean Lorentz factor measured in a frame that moves with the local $\boldsymbol{v}_{\rm E\times B}=c(\boldsymbol{E}\times \boldsymbol{B})/B^2$ (hereafter, the $E\times B$ frame). In the same frame, we measure the work done by $B-$parallel electric fields (red lines in \figg{line}) as
\begin{equation}
(\langle\gamma_{\rm e,E_\parallel}\rangle-\gamma_{\rm e1}) m_{\rm e} c^2=\left\langle\int -e E_\parallel v_{\parallel} dt \right\rangle
\end{equation}
where $\langle \cdot\rangle $ stands for an average over the electrons we are tracking, and $v_{\parallel}$ is the $B-$parallel 3-velocity of an individual electron in the local $E\times B$ frame, where $E_\parallel$ is also computed. The work done by magnetic field compression, assuming conservation of the adiabatic invariants $\gamma\beta_\parallel$ and $(\gamma\beta_\perp)^2/B$, can be computed as follows. First, the change in Lorentz factor
for each electron between timestep $n$ and $n+1$ is calculated as in \citet{tran_20},
\begin{equation}
\gamma_{n\rightarrow n+1,\rm
 B}=\sqrt{1+(\gamma\beta_\parallel)^2_{n}+(\gamma\beta_\perp)^2_{n}\, (B_{n+1}/B_n})
\end{equation}
where $\beta_\parallel=v_\parallel/c$, while $\beta_\perp$ is the dimensionless electron velocity perpendicular to the local magnetic field, both measured in the $E\times B$ frame. The electron Lorentz factor in the $E\times B$ frame is $\gamma=1/\sqrt{1-\beta_\parallel^2-\beta_\perp^2}$. When averaged over the  population of tracked electrons (blue solid lines in \figg{line}), we have
\begin{equation}
\langle\gamma_{\rm e,B}\rangle-\gamma_{\rm e1}=\left\langle \Sigma_n (\gamma_{n\rightarrow n+1,\rm
 B}-\gamma_{n})\right\rangle~.
\end{equation}
The compressive contribution $\langle\gamma_{\rm e,B}\rangle$ can be compared with the adiabatic expectation for a 3D isotropic ultra-relativistic gas (dashed blue lines in \figg{line})
\begin{equation}
\langle\gamma_{\rm e,ad3D}\rangle=\langle(n/n_{1})^{1/3} \gamma_{\rm 1}\rangle
\end{equation}
(where the density $n$ is measured in the local $E\times B$ frame, and $\gamma_{\rm 1}$ is the pre-shock Lorentz factor of an individual electron, such that $\gamma_{\rm e1}=\langle\gamma_1\rangle$), or with the expectation for a 2D fluid that preserves the adiabatic invariants since the beginning (dotted blue lines in \figg{line})
\begin{equation}
\langle\gamma_{\rm e,ad2D}\rangle=\left\langle\sqrt{1+(\gamma\beta_\parallel)^2_{1}+(\gamma\beta_\perp)^2_{1}\, (B/B_1})\right\rangle
\end{equation}
where the subscript $1$ indicates initial conditions of each electron (i.e., at the selection time $t_{\rm sel}$). If the adiabatic invariance were to be always preserved, $\langle\gamma_{\rm e,B}\rangle=\langle\gamma_{\rm e,ad2D}\rangle$.

We remark that super-adiabatic heating via magnetic pumping is quantified by the difference $\langle\gamma_{\rm e,B}\rangle-\langle\gamma_{\rm e,ad3D}\rangle$ (i.e., the difference between solid and dashed blue lines in \figg{line}). Also, the overall amount of super-adiabatic heating $\te-\tead$ discussed before is proportional to the difference between the black solid line and  blue  dashed line at late times (in the ultra-relativistic limit, $\theta_{\rm e}= \gamma_{\rm e}/3$).

In \figg{line}, the dotted black lines illustrate the combined contributions of $B-$parallel heating and magnetic compression, showing that their sum is a good proxy for the overall heating curve (black solid lines), for both $\tratio=1$ (top) and $\tratio=0.1$ (bottom). We now comment on the trends established after the heating curves have reached a nearly constant value, i.e., $\Omega_{\rm i}(t-t_{\rm sel})\gtrsim 8$. Standard adiabatic compression ($\langle\gamma_{\rm e,ad3D}\rangle$, dashed blue lines) accounts for $\sim 50\%$ of the overall heating at $\tratio=1$ and for  $\sim 25\%$ at $\tratio=0.1$. For both $\tratio=1$ and $\tratio=0.1$, irreversible heating by magnetic pumping (i.e., the difference between solid and dashed blue lines) amounts to $\sim 50\%$ of the 3D adiabatic expectation. Heating by $B-$parallel electric fields contributes $\sim 25\%$ of the overall heating  at $\tratio=1$ and $\sim 50\%$  at $\tratio=0.1$.

At $\Omega_{\rm i}(t-t_{\rm sel})\gtrsim 8$, heating by magnetic compression (solid blue lines) scales such that $\langle\gamma_{\rm e,B}\rangle/\gamma_{\rm e1}-1$ is roughly independent of $\tratio$. It follows that the main reason why the overall $\langle\gamma_{\rm e}\rangle/\gamma_{\rm e1}-1$ is larger for $\tratio=0.1$ than for $\tratio=1$ (see also \figg{eff}(a)) is the additional contribution of $B-$parallel electric field work.

The same conclusions can be extracted from \figg{all}, where we present the contributions of various heating mechanisms to the far-downstream electron mean energy, as a function of $\bp$ and $\tratio$. We find that, in most cases, the sum of $B-$parallel heating and magnetic compression can account for the overall electron energy change. As regard to super-adiabatic heating, magnetic pumping dominates for higher $\tratio$ and, at fixed $\tratio$, it increases with $\bp$ (it amounts to a fraction $\sim 50\%$  of the 3D adiabatic expectation at $\bp=1$ and $\sim 100\%$ at $\bp=8$). In contrast, irreversible heating by $B-$parallel electric fields dominates for $\tratio=0.1$.
As we have already remarked, heating by magnetic compression scales such that $\langle\gamma_{\rm e,B}\rangle-\gamma_{\rm e1}\propto \gamma_{\rm e1} $, at each fixed $\bp$. In contrast, the contribution $\langle\gamma_{\rm e,E_\parallel}\rangle-\gamma_{\rm e1}$ by $B-$parallel electric field work has a shallower scaling, since it increases by less than a factor of three between $\tratio=0.1$ and $\tratio=1$.

\section{Discussion and Conclusions}
\label{sec:discuss}
In this paper, we have used 2D PIC simulations to quantify the efficiency and mechanisms of electron heating at the collisionless shocks detected in GRMHD simulations of tilted accretion disks. For geometrically-thick, radiatively inefficient accretion flows, these shocks have trans-relativistic speed, moderate plasma beta, and low sonic Mach number --- a parameter regime still largely unexplored. We find that  the post-shock electron temperature $T_{\rm e2}$ exceeds the adiabatic expectation $T_{\rm e2, ad}$ by approximately
$\htgequation$, nearly independent of the plasma beta and of the temperature ratio. This approximation may be used to incorporate the efficiency of shock-driven electron heating into GRMHD simulations of tilted accretion disks. We also investigate the mechanisms of electron heating, and find that for $\Msim=5$ (i.e., $M_s\sim5-6$) it is governed by magnetic pumping at $\tez/\tiz=1$, while heating by $B-$parallel electric fields dominates at $\tez/\tiz=0.1$.

Our results have been obtained for strictly perpendicular shocks. We expect that our conclusions will also apply to quasi-perpendicular superluminal shocks, while different outcomes may be expected for quasi-perpendicular subluminal shocks, where shock-reflected electrons can propagate back upstream (for a study of electron heating in non-relativistic quasi-perpendicular shocks, see \citealt{tran_23}).
In quasi-parallel shocks, protons can be efficiently reflected back upstream and accelerated via the Fermi process, and the electron heating physics is likely to be strongly affected by the properties of non-thermal protons and their self-generated waves. Such an investigation will be the subject of future work.

\begin{acknowledgments}
Conversations with Jordy Davelaar, Jason Dexter, Francisco Ley, Matthew Liska, Nick Kaaz and Ellen Zweibel are gratefully acknowledged.
AT and LS were partly supported by NASA ATP 80NSSC20K0565.
AT was partly supported by NASA FINESST 80NSSC21K1383 and NSF PHY-2010189. LS acknowledges support from  DoE Early Career Award DE-SC0023015. This work was supported by a grant from the Simons Foundation (MP-SCMPS-00001470) to LS, and facilitated by Multimessenger Plasma Physics Center (MPPC), NSF grant PHY-2206609.
We are indebted to Xinyi Guo for analysis code.
This work was expedited by NASA's Astrophysics Data System, Jonathan Sick's and
Rui Xue's \texttt{ads2bibdesk}, Benty Fields, and the Python / Matplotlib /
Numpy stack. LS is grateful for hospitality to the KITP, which is funded by NSF PHY-1748958.
\end{acknowledgments}

\bibliographystyle{aasjournal}
\bibliography{library,lorenzo_bib}

\appendix

\section{Simulation Parameters} \label{app:param}

Table~\ref{tab:param} provides simulation input parameters, defined as follows.
\begin{itemize}
    \item $\Msim$ is the simulation-frame ion-sound Mach number (Section~\ref{sec:methods}).
    \item $M_s$ is the measured sonic Mach number (Section~\ref{sec:methods}).
    \item $\beta_\mt{pi1}$ is the upstream ion plasma beta (Section~\ref{sec:methods}).
    \item $\tez/\tiz$ is the upstream electron/ion temperature ratio
        (Section~\ref{sec:methods}).
    \item $v_0/c$ is the simulation-frame upstream plasma flow speed
        (Section~\ref{sec:methods}).
    \item $v_1/c$ is the measured shock speed in the upstream frame
        (Section~\ref{sec:methods}).
    \item $\Omci t$ is the simulation time shown in Section~\ref{sec:results}
        and used to measure $x_\mt{xshock}$ (Section~\ref{sec:methods});
        it is also equal to the selection time $\Omci t_{\rm sel}$ for
        particle-tracing analysis in Section~\ref{sec:heating}.
    \item $x_\mathrm{shock}$ is the shock location at $t\Omci$ in units of
        $\rho_\mathrm{i}$.
    \item $x_\mathrm{L}$ and $x_\mathrm{R}$ define the interval wherein the
        downstream flow temperatures $\te$ and $\ti$ are measured
        (Section~\ref{sec:heating}).
        Both $x_\mathrm{L}$ and $x_\mathrm{R}$ are defined as offsets from
        $x_\mathrm{shock}$;
        both $x_\mathrm{L}$ and $x_\mathrm{R}$ are reported in units of
        $\rho_\mathrm{i}$.
    \item $\te/\tead-1$, $\te/T_{\rm e1}$, $\ti/T_{\rm i1}$, and $\te/\ti$
        quantify the post-shock ion and electron thermal energy gain, measured
        as a volume average over the spatial interval
        $x - x_\mathrm{shock} \in [x_\mathrm{L}, x_\mathrm{R}]$
        (Section~\ref{sec:heating}).
\end{itemize}

\begin{deluxetable*}{llll ll rrrr rrrr}
\tabletypesize{\scriptsize}
\tablecaption{
    Simulation input parameters.
    Columns are defined in Sections~\ref{sec:methods} and \ref{sec:heating}, and
    Appendix~\ref{app:param}.
    \label{tab:param}
}
\tablehead{
    $\Msim$
    & $M_s$
    & $\beta_\mt{pi1}$
    & $\tez/\tiz$
    & $v_0/c$
    & $v_1/c$
    & $\Omega_\mt{i} t$
    & $x_\mt{shock}$
    & $x_\mt{L}$
    & $x_\mt{R}$
    & $\te/\tead-1$
    & $\te/T_{\rm e1}$
    & $\Ti/T_{\rm i1}$
    & $\te/\ti$
}
\startdata
\input{parameters_formatted.tex}\enddata  
\tablecomments{
Table~\ref{tab:param} is
available in a machine-readable format in the online journal.
}
\end{deluxetable*}

\end{document}

%% file: parameters_formatted.tex
2.14 & 3.45 & 1.03 & 0.10 & 0.276 & 0.456 & 40.66 & 29.1 & $-$20.0 & $-$10.0 & 0.10 & 1.485 & 3.282 & 0.045 \\
2.14 & 3.24 & 1.03 & 0.32 & 0.276 & 0.459 & 40.66 & 29.7 & $-$20.0 & $-$10.0 & 0.07 & 1.432 & 3.213 & 0.141 \\
2.14 & 2.79 & 1.03 & 1.00 & 0.276 & 0.469 & 40.66 & 31.3 & $-$20.0 & $-$10.0 & 0.05 & 1.389 & 3.076 & 0.452 \\
\hline
3.21 & 4.41 & 1.00 & 0.10 & 0.414 & 0.583 & 40.13 & 19.6 & $-$14.0 & $-$6.0 & 0.24 & 1.820 & 6.429 & 0.028 \\
3.21 & 4.12 & 1.00 & 0.32 & 0.414 & 0.585 & 40.13 & 19.8 & $-$14.0 & $-$6.0 & 0.16 & 1.685 & 6.349 & 0.084 \\
3.21 & 3.53 & 1.00 & 1.00 & 0.414 & 0.593 & 40.13 & 20.9 & $-$14.0 & $-$6.0 & 0.10 & 1.597 & 6.094 & 0.262 \\
\hline
4.28 & 5.32 & 0.96 & 0.10 & 0.552 & 0.703 & 40.17 & 15.0 & $-$9.0 & $-$4.0 & 0.61 & 2.529 & 11.369 & 0.022 \\
4.28 & 4.96 & 0.96 & 0.32 & 0.552 & 0.704 & 40.17 & 15.0 & $-$9.0 & $-$4.0 & 0.34 & 2.094 & 11.174 & 0.059 \\
4.28 & 4.21 & 0.96 & 1.00 & 0.552 & 0.707 & 40.17 & 15.4 & $-$9.0 & $-$4.0 & 0.22 & 1.913 & 10.773 & 0.178 \\
\hline
5.35 & 6.15 & 0.90 & 0.10 & 0.690 & 0.813 & 40.32 & 11.8 & $-$8.0 & $-$2.0 & 1.50 & 4.197 & 19.901 & 0.021 \\
5.35 & 5.74 & 0.90 & 0.32 & 0.690 & 0.814 & 40.32 & 11.9 & $-$8.0 & $-$2.0 & 0.69 & 2.837 & 19.827 & 0.045 \\
5.35 & 4.85 & 0.90 & 1.00 & 0.690 & 0.814 & 40.32 & 12.0 & $-$8.0 & $-$2.0 & 0.43 & 2.389 & 19.323 & 0.124 \\
\hline\hline
2.14 & 3.32 & 2.05 & 0.10 & 0.276 & 0.439 & 25.16 & 16.2 & $-$13.0 & $-$8.0 & 0.12 & 1.532 & 3.531 & 0.043 \\
2.14 & 3.12 & 2.05 & 0.32 & 0.276 & 0.442 & 25.16 & 16.6 & $-$13.0 & $-$8.0 & 0.08 & 1.476 & 3.446 & 0.135 \\
2.14 & 2.68 & 2.05 & 1.00 & 0.276 & 0.451 & 25.16 & 17.5 & $-$13.0 & $-$8.0 & 0.06 & 1.430 & 3.286 & 0.435 \\
\hline
3.21 & 4.31 & 2.00 & 0.10 & 0.414 & 0.571 & 25.99 & 11.7 & $-$8.0 & $-$3.0 & 0.26 & 1.860 & 6.796 & 0.027 \\
3.21 & 4.04 & 2.00 & 0.32 & 0.414 & 0.573 & 25.99 & 11.9 & $-$8.0 & $-$3.0 & 0.18 & 1.752 & 6.687 & 0.083 \\
3.21 & 3.45 & 2.00 & 1.00 & 0.414 & 0.579 & 25.99 & 12.4 & $-$8.0 & $-$3.0 & 0.13 & 1.653 & 6.455 & 0.256 \\
\hline
4.28 & 5.24 & 1.92 & 0.10 & 0.552 & 0.692 & 25.62 & 8.8 & $-$5.0 & $-$2.5 & 0.62 & 2.582 & 11.875 & 0.022 \\
4.28 & 4.88 & 1.92 & 0.32 & 0.552 & 0.693 & 25.62 & 8.8 & $-$5.0 & $-$2.5 & 0.37 & 2.183 & 11.766 & 0.059 \\
4.28 & 4.15 & 1.92 & 1.00 & 0.552 & 0.697 & 25.62 & 9.1 & $-$5.0 & $-$2.5 & 0.25 & 1.978 & 11.401 & 0.174 \\
\hline
5.35 & 6.08 & 1.80 & 0.10 & 0.690 & 0.804 & 25.45 & 6.9 & $-$4.0 & $-$1.0 & 1.21 & 3.775 & 20.591 & 0.018 \\
5.35 & 5.67 & 1.80 & 0.32 & 0.690 & 0.804 & 25.45 & 6.8 & $-$4.0 & $-$1.0 & 0.70 & 2.901 & 20.264 & 0.045 \\
5.35 & 4.79 & 1.80 & 1.00 & 0.690 & 0.805 & 25.45 & 6.9 & $-$4.0 & $-$1.0 & 0.45 & 2.479 & 19.676 & 0.126 \\
\hline\hline
2.14 & 3.24 & 4.11 & 0.10 & 0.276 & 0.429 & 25.77 & 15.5 & $-$12.0 & $-$5.0 & 0.11 & 1.548 & 3.755 & 0.041 \\
2.14 & 3.05 & 4.11 & 0.32 & 0.276 & 0.433 & 25.77 & 16.0 & $-$12.0 & $-$5.0 & 0.08 & 1.492 & 3.648 & 0.129 \\
2.14 & 2.62 & 4.11 & 1.00 & 0.276 & 0.440 & 25.77 & 16.8 & $-$12.0 & $-$5.0 & 0.07 & 1.451 & 3.430 & 0.423 \\
\hline
3.21 & 4.21 & 4.00 & 0.10 & 0.414 & 0.556 & 25.15 & 10.2 & $-$7.0 & $-$2.5 & 0.30 & 1.985 & 7.045 & 0.028 \\
3.21 & 3.94 & 4.00 & 0.32 & 0.414 & 0.559 & 25.15 & 10.4 & $-$7.0 & $-$2.5 & 0.23 & 1.864 & 6.863 & 0.086 \\
3.21 & 3.36 & 4.00 & 1.00 & 0.414 & 0.563 & 25.15 & 10.8 & $-$7.0 & $-$2.5 & 0.15 & 1.733 & 6.520 & 0.266 \\
\hline
4.28 & 5.16 & 3.84 & 0.10 & 0.552 & 0.682 & 25.99 & 8.2 & $-$5.5 & $-$1.5 & 0.51 & 2.460 & 12.202 & 0.020 \\
4.28 & 4.81 & 3.84 & 0.32 & 0.552 & 0.682 & 25.99 & 8.2 & $-$5.5 & $-$1.5 & 0.39 & 2.273 & 12.088 & 0.059 \\
4.28 & 4.07 & 3.84 & 1.00 & 0.552 & 0.683 & 25.99 & 8.3 & $-$5.5 & $-$1.5 & 0.26 & 2.049 & 11.456 & 0.179 \\
\hline
5.35 & 6.02 & 3.61 & 0.10 & 0.690 & 0.797 & 25.20 & 6.2 & $-$4.0 & $-$1.5 & 1.30 & 4.020 & 21.084 & 0.019 \\
5.35 & 5.62 & 3.61 & 0.32 & 0.690 & 0.797 & 25.20 & 6.3 & $-$4.0 & $-$1.5 & 0.72 & 3.001 & 20.813 & 0.046 \\
5.35 & 4.76 & 3.61 & 1.00 & 0.690 & 0.799 & 25.20 & 6.4 & $-$4.0 & $-$1.5 & 0.45 & 2.513 & 20.440 & 0.123 \\
\hline\hline
2.14 & 3.17 & 8.21 & 0.10 & 0.276 & 0.419 & 25.16 & 14.1 & $-$10.0 & $-$5.0 & 0.14 & 1.604 & 3.977 & 0.040 \\
2.14 & 2.98 & 8.21 & 0.32 & 0.276 & 0.423 & 25.16 & 14.6 & $-$10.0 & $-$5.0 & 0.09 & 1.533 & 3.854 & 0.126 \\
2.14 & 2.56 & 8.21 & 1.00 & 0.276 & 0.430 & 25.16 & 15.3 & $-$10.0 & $-$5.0 & 0.07 & 1.482 & 3.587 & 0.413 \\
\hline
3.21 & 4.13 & 8.00 & 0.10 & 0.414 & 0.546 & 25.99 & 9.7 & $-$7.0 & $-$2.0 & 0.29 & 2.008 & 7.152 & 0.028 \\
3.21 & 3.86 & 8.00 & 0.32 & 0.414 & 0.547 & 25.99 & 9.8 & $-$7.0 & $-$2.0 & 0.22 & 1.903 & 6.983 & 0.086 \\
3.21 & 3.31 & 8.00 & 1.00 & 0.414 & 0.555 & 25.99 & 10.5 & $-$7.0 & $-$2.0 & 0.14 & 1.741 & 6.599 & 0.264 \\
\hline
4.28 & 5.09 & 7.68 & 0.10 & 0.552 & 0.673 & 25.17 & 7.3 & $-$5.5 & $-$1.5 & 0.62 & 2.687 & 12.492 & 0.022 \\
4.28 & 4.76 & 7.68 & 0.32 & 0.552 & 0.675 & 25.17 & 7.5 & $-$5.5 & $-$1.5 & 0.38 & 2.292 & 12.322 & 0.059 \\
4.28 & 4.05 & 7.68 & 1.00 & 0.552 & 0.679 & 25.17 & 7.7 & $-$5.5 & $-$1.5 & 0.24 & 2.041 & 11.852 & 0.172 \\
\hline
5.35 & 6.00 & 7.22 & 0.10 & 0.690 & 0.793 & 25.81 & 6.1 & $-$5.0 & $-$1.0 & 1.42 & 4.253 & 21.312 & 0.020 \\
5.35 & 5.58 & 7.22 & 0.32 & 0.690 & 0.792 & 25.81 & 6.1 & $-$5.0 & $-$1.0 & 0.77 & 3.123 & 21.103 & 0.047 \\
5.35 & 4.73 & 7.22 & 1.00 & 0.690 & 0.793 & 25.81 & 6.2 & $-$5.0 & $-$1.0 & 0.51 & 2.654 & 20.572 & 0.129